\documentclass[twocolumn,aps,prl,showpacs]{revtex4}

\usepackage{amsmath}

\begin{document}
\title{Bipartite Bell Inequalities for Hyperentangled States}
\author{Ad\'{a}n Cabello}
\email{adan@us.es}
\affiliation{Departamento de F\'{\i}sica Aplicada II,
Universidad de Sevilla, 41012 Sevilla, Spain}
\date{\today}


\begin{abstract}
We show that bipartite Bell inequalities based on the
Einstein-Podolsky-Rosen criterion for elements of reality and
derived from the properties of some hyperentangled states allow
feasible experimental verifications of the fact that quantum
nonlocality grows exponentially with the size of the subsystems, and
Bell loophole-free tests with currently available photodetection
efficiencies.
\end{abstract}


\pacs{03.65.Ud,
03.67.Pp}

\maketitle




Bell's theorem states that quantum mechanics cannot be reproduced by
any local realistic theory \cite{Bell64}. Therefore, either ``there
must be a mechanism whereby the setting of one measuring device can
influence the reading of another instrument, however remote''
\cite{Bell64}, or we must give up on the idea that some physical
observables possess definite values. This result is usually referred
to as ``quantum nonlocality.'' The violation of a Bell inequality
(BI) is a standard method used to identify quantum nonlocality.

A BI is a constraint imposed by local realistic theories on the
values of a linear combination $\hat{\beta}$ of the averages (or
probabilities) of the results of experiments on two or more
separated systems. It takes the form $\hat{\beta} \le \beta$, where
the bound $\beta$ is the maximal possible value of $\hat{\beta}$
allowed by the local realistic theories. There are two types of BIs
depending on how we define ``local realistic theories.''

{\em CHSH-BIs.}---The most common BIs belong to the
Clauser-Horne-Shimony-Holt (CHSH) type \cite{CHSH69,CH74}, in which
local realistic theories are defined as those in which: (i) the
probabilities of the outcomes of {\em all} local observables are
predetermined, and (ii) these probabilities cannot be affected by
spacelike separated measurements. For two separated systems $1$ and
$2$, in any CHSH-BI, $\hat{\beta}$ takes the following general form
\begin{equation}
\hat{\beta} = \sum_{i=1}^m
\sum_{j=1}^n c(i,j) \langle A_1^{(i)} B_2^{(j)} \rangle,
\label{BellCHSHoperator}
\end{equation}
where $c(i,j)$ are certain
constant coefficients, $i$ and $j$ are indices (discrete or
continuous) distinguishing the possible experiments on system $1$
and $2$, respectively, and $\langle A_1^{(i)} B_2^{(j)} \rangle$ is
a correlation function (the average of the product of the
observables measured on $1$ and $2$). If $A_1^{(i)}$ and $B_2^{(j)}$
are spacelike separated experiments, from assumptions (i) and (ii)
follow that the correlation must take the form
\begin{equation}
\langle A_1^{(i)} B_2^{(j)} \rangle = \sum_{\lambda} f_1 (A_1^{(i)},
\lambda) g_2 (B_2^{(j)}, \lambda) p (\lambda),
\label{Deterministiclocaltheories}
\end{equation}
where $f_1(A_1^{(i)}, \lambda)$ [$g_2 (B_2^{(j)}, \lambda)$] is a
function which gives the value of the experiment $A_1^{(i)}$
($B_2^{(j)}$) on subsystem $1$ ($2$), and $\lambda$ is a summation
(or integration) parameter which allows the description to have a
probabilistic nature; $p(\lambda)$ is the distribution of the
parameter.

Curiously, the original BI \cite{Bell64} does {\em not} belong to
this type. Any product state satisfies any CHSH-BI. Therefore,
separable states (which are convex combinations of product states)
satisfy them too. However, some separable states violate the
original BI \cite{Loubenets04}. The explanation is that the original
BI is based on assumptions that are not satisfied by these separable
states \cite{Redhead87}.

{\em EPR-BIs.}---The premise of the original BI is the
Einstein-Podolsky-Rosen (EPR) criterion for the existence of
elements of reality: ``if, without in any way disturbing a system,
we can predict with certainty (i.e., with probability equal to
unity) the value of a physical quantity, then there exists an
element of physical reality corresponding to this physical
quantity'' \cite{EPR35}. The EPR criterion establishes {\em two}
conditions for the existence of elements of reality. Firstly, {\em
perfect predictability}: it must be possible to predict them with
certainty. Secondly, {\em locality}: the prediction must be based
on a measurement that exerts no disturbing influence upon them.

Bell's original inequality is based on an {\em equality}: it is
based on the fact that, for the two-qubit singlet state, the results
of measuring the same observable $B$ on both qubits are perfectly
anticorrelated, $\langle B B \rangle = -1$. Bell uses this equality
in two ways: to guarantee that all local observables are EPR
elements of reality, and in the derivation of the inequality.




What makes EPR-BIs so attractive is that the EPR criterion seems
almost unavoidable: EPR do {\em not} assume that all local
experiments should have predefined values; the existence of
predefined values is assumed to explain why they can be predicted
from remote measurements. On the other hand, the advantage of the
CHSH-BIs is that they do not depend on the properties of a
particular state. The problem for testing EPR-BIs is the difficulty
of having perfect correlations in actual experiments.

The work by Greenberger, Horne, and Zeilinger (GHZ) \cite{GHZ89},
and Mermin \cite{Mermin90a} stimulated a renewed interest in the
local realistic theories of the EPR type. Moreover, the development
of quantum technologies in the last decade has opened new
possibilities.


{\em Almost perfect predictability.}---The first interesting
development is that we can prepare two-particle states with almost
perfect correlations. This almost perfect remote predictability
opens the door for the testing of the EPR-BIs. There are two
possible strategies: One is to relax the EPR criterion and define
elements of reality as those that can be predicted with {\em almost}
perfect certainty \cite{ER95}. This definition would automatically
extend the validity of the BI to all prepared pairs. The second
strategy is valid if we can prepare pairs so that a very high
fraction of them allows perfect predictability. Then we can assume
the original EPR criterion. In this case, the EPR-BI is legitimate
only for a fraction of pairs. However, we still can obtain
conclusive experimental results by performing tests on all the
pairs, and then calculating how the fraction of the pairs for which
the inequality is not valid can affect these experimental results.


{\em Hyperentanglement and detection efficiency.}---The second
interesting development is the possibility of preparing pairs of
particles in hyperentangled states, i.e. entangled in several
degrees of freedom \cite{Kwiat97}. Hyperentanglement has been
demonstrated in recent experiments with two photons entangled in two
degrees of freedom (polarization and path) \cite{CBPMD05}, and in
three degrees of freedom (polarization, path, and time-energy)
\cite{BLPK05}. Indeed, time-bin entanglement would allow us to
encapsulate a higher number of qubits \cite{Gisin06}. This is very
interesting for the following reason: imagine we have $2N$ qubits
distributed in $2N$ particles; then, to ``reveal'' $2N$ EPR elements
of reality we would need to activate $2N$ single-particle detectors,
something that occurs with probability $\eta^{2N}$, being $\eta$ the
efficiency of each of the single-particle detectors. However, if we
have $2N$ qubits encapsulated in $2$ particles; then, to reveal $2N$
EPR elements of reality we would only need to activate $2$
single-particle detectors, something that occurs with probability
$\eta^{2}$. The interest of this is related to the fact that the
main obstacle for a loophole-free test of BIs is that $\eta$ is very
low for photons.


{\em Simultaneous EPR elements of reality.}---A third motivation for
exploring EPR-BIs is related with the experimental capability of
entangling higher dimensional subsystems. While the $\hat{\beta}$
corresponding to bipartite CHSH-BIs takes the same form
(\ref{BellCHSHoperator}) irrespective of the dimension $d$ of the
Hilbert space describing the local subsystems, the $\hat{\beta}$
corresponding to bipartite EPR-BIs can take different forms
depending on $d$. The interesting point is that if $d > 2$, there
are {\em compatible} local observables which can be regarded as {\em
simultaneous} EPR elements of reality \cite{HR83}. For instance,
suppose that $A_1^{(1)}$ and $A_1^{(2)}$ are observables on particle
$1$ represented by {\em commuting} operators. Now there is a new
possibility: it can so happen that there are a quantum state
$|\phi\rangle$ and two local observables $B_2^{(1)}$ and
$B_2^{(2)}$, on particle $2$ so that $\langle \phi |A_1^{(1)}
B_2^{(1)}|\phi \rangle= 1$ and $\langle \phi |A_1^{(2)}
B_2^{(2)}|\phi \rangle= 1$. Therefore, since $A_1^{(1)}$ and
$A_1^{(2)}$ can be remotely predicted with certainty, then {\em
both} $A_1^{(1)}$ and $A_1^{(2)}$ are EPR elements of reality {\em
simultaneously}. In principle, it could happen that, since
$A_1^{(1)}$ and $A_1^{(2)}$ are measured on the same subsystem, the
measurement of one of them may disturb the value of the other. The
remarkable point is that this hypothetical disturbance can be
discarded if a spacelike separated observer {\em can} predict with
certainty the values of $A_1^{(1)}$ or $A_1^{(2)}$, not only when
they are measured separately, but also when they are measured
together.

Therefore, for higher dimensional bipartite EPR-BIs, $\hat{\beta}$
can contain terms like $\langle A_1^{(1)} A_1^{(2)} B_2^{(1)}
\rangle$, and the general form of $\hat{\beta}$ is
\begin{eqnarray}
\hat{\beta} & = & \sum_{i=1}^m \ldots
\sum_{j=1}^n \sum_{k=1}^p \ldots \sum_{l=1}^q c(i, \ldots, j, k,
\ldots, l) \nonumber \\ & & \times \langle A_1^{(i)} \ldots
A_1^{(j)} B_2^{(k)} \ldots B_2^{(l)} \rangle,
\label{BellEPRoperator}
\end{eqnarray}
where all of the local observables are EPR elements of reality (for
certain states), and all the local observables appearing in the same
average are compatible. Examples of bipartite EPR-BIs have been
introduced \cite{Cabello01b,Aravind02,Cabello05} and experimentally
tested \cite{CBPMD05} recently.

The aim of this Letter is to explore the merits of these new
equalities-based BIs when we move to higher dimensions. For this
purpose, we derive a bipartite higher dimensional EPR-BI based on
the properties of a hyperentangled state, and show that it can be
used to solve two still-open experimental problems in quantum
mechanics.




{\em Growing with size nonlocality.}---There was a time when it was
thought that quantum nonlocality would decrease as the size of the
system grows, as a manifestation of some intrinsic aspect of the
transition from quantum to classical behavior. By ``size'' we mean
either the number of particles or the number of internal degrees of
freedom. However, Mermin \cite{Mermin90b} showed that the
correlations found by $n$ spacelike separated observers that share
$n$ qubits in a GHZ state violate a $n$-party (with $n \ge 3$) BI by
a factor that increases {\em exponentially} with $n$. An
experimental verification of this exponentially growing nonlocality
using GHZ states is difficult because it requires $n$ spacelike
separated measurements, and because $n$-party GHZ states'
sensitivity to decoherence also grows with $n$ \cite{HDB05}.

The ratio $\beta_{\rm EXP} / \beta_{\rm EPR}$, where $\beta_{\rm
EXP}$ is the experimental value of $\hat{\beta}$ (which is
supposedly similar to $\beta_{\rm QM}$), and $\beta_{\rm EPR}$ is
the maximal possible value of $\hat{\beta}$ allowed by the local
realistic theories of the EPR-type, is a good measure of
nonlocality, since it is related both to the number of bits needed
to communicate nonlocally in order to emulate the experimental
results by a local realistic theory, and also to the minimum
detection efficiency needed for a loophole-free experiment (as
explained below). In all known bipartite CHSH-BIs this ratio is
almost {\em constant} with the number of internal levels of the
local subsystems \cite{Gisin06,GP92}. However, this is not the case
in the following EPR-BI.

Consider two particles $1$ and $2$ prepared in the state
\begin{equation}
|\Psi\rangle = \bigotimes_{j=1}^N
|\psi\rangle^{(j)},
\end{equation}
where
\begin{eqnarray}
|\psi\rangle^{(j)} & = & \frac{1}{2} \left(|00\rangle^{(j)}_1
|00\rangle^{(j)}_2+ |01\rangle^{(j)}_1 |01\rangle^{(j)}_2 +
|10\rangle^{(j)}_1 |10\rangle^{(j)}_2 \right. \nonumber \\ & &
\left. - |11\rangle^{(j)}_1 |11\rangle^{(j)}_2\right).
\label{benasque}
\end{eqnarray}
The state $|\Psi\rangle$ encapsulates $4N$ qubits in two particles.
Consider the following single qubit observables:
\begin{align}
X_k^{(j)} = \sigma_x^{(j)} \otimes 1\!\!\:\!{\rm{I}}^{(j)},\; &
Y_k^{(j)} = \sigma_y^{(j)} \otimes 1\!\!\:\!{\rm{I}}^{(j)},\;
Z_k^{(j)} = \sigma_z^{(j)} \otimes 1\!\!\:\!{\rm{I}}^{(j)}, \\
x_1^{(j)} = 1\!\!\:\!{\rm{I}}^{(j)} \otimes \sigma_x^{(j)},\; &
y_2^{(j)} = 1\!\!\:\!{\rm{I}}^{(j)} \otimes \sigma_y^{(j)},\;
z_2^{(j)} = 1\!\!\:\!{\rm{I}}^{(j)} \otimes \sigma_z^{(j)},
\end{align}
where $k$ denotes particle $k$, $\sigma_x$ is the Pauli matrix in
the $x$ direction, and $1\!\!\:\!{\rm{I}}$ is the identity matrix in
a two-dimensional Hilbert space. For the state $|\Psi\rangle$, each
and every one of these $7N$ single qubit observables $X_1^{(j)}$,
$Y_1^{(j)}$, $x_1^{(j)}$, $X_2^{(j)}$, $Y_2^{(j)}$, $y_2^{(j)}$, and
$z_2^{(j)}$ can be regarded as an EPR element of reality, since it
satisfies the following $7N$ equalities representing perfect
correlations:
\begin{eqnarray}
\langle X_1^{(j)} X_2^{(j)} z_2^{(j)} \rangle = 1,\;\;\;\;
\langle Y_1^{(j)} Y_2^{(j)} z_2^{(j)} \rangle & = & -1,\\
\langle x_1^{(j)} Z_2^{(j)} x_2^{(j)} \rangle & = & 1,\\
\langle X_1^{(j)} z_1^{(j)} X_2^{(j)} \rangle = 1,\;\;\;\;
\langle Y_1^{(j)} z_1^{(j)} Y_2^{(j)} \rangle & = & -1,\\
\langle Z_1^{(j)} y_1^{(j)} y_2^{(j)} \rangle = -1,\;\;\;\;
\langle z_1^{(j)} z_2^{(j)} \rangle & = & 1.
\end{eqnarray}
Therefore, we can define
\begin{widetext}
\begin{eqnarray}
\beta & = &
\langle X_1^{(1)} X_2^{(1)} z_2^{(1)} \ldots X_1^{(N-1)} X_2^{(N-1)} z_2^{(N-1)} X_1^{(N)} X_2^{(N)} z_2^{(N)} \rangle
\nonumber \\ & &
- \langle X_1^{(1)} X_2^{(1)} z_2^{(1)} \ldots X_1^{(N-1)} X_2^{(N-1)} z_2^{(N-1)} Y_1^{(N)} Y_2^{(N)} z_2^{(N)} \rangle
\nonumber \\ & &
+ \langle X_1^{(1)} X_2^{(1)} z_2^{(1)} \ldots X_1^{(N-1)} X_2^{(N-1)} z_2^{(N-1)} X_1^{(N)} x_1^{(N)} Y_2^{(N)} y_2^{(N)} \rangle
\nonumber \\ & &
+ \langle X_1^{(1)} X_2^{(1)} z_2^{(1)} \ldots X_1^{(N-1)} X_2^{(N-1)} z_2^{(N-1)} Y_1^{(N)} x_1^{(N)} X_2^{(N)} y_2^{(N)} \rangle
\nonumber \\ & &
- \langle X_1^{(1)} X_2^{(1)} z_2^{(1)} \ldots Y_1^{(N-1)} Y_2^{(N-1)} z_2^{(N-1)} X_1^{(N)} X_2^{(N)} z_2^{(N)} \rangle
+ \ldots
\nonumber \\ & &
+ \langle Y_1^{(1)} x_1^{(1)} X_2^{(1)} y_2^{(1)} \ldots Y_1^{(N)} x_1^{(N)} X_2^{(N)} y_2^{(N)} \rangle,
\label{Belloperator}
\end{eqnarray}
\end{widetext}
which contains $4^N$ expectation values. For measuring, for
instance, $X_1^{(1)} x_1^{(1)} \ldots X_1^{(N)} x_1^{(N)}$ on
particle $1$, we use an analyzer that separates the two
possibilities of each of the $2N$ qubit observables $X_1^{(1)}$,
$x_1^{(1)}$, \ldots, $X_1^{(N)}$, $x_1^{(N)}$. This analyzer is
backed up by $4^N$ particle detectors, one for each of the possible
outcomes. Therefore, each particle detection gives the value of $2N$
observables. Each observer can choose between the $4^N$ local
experiments. The choice of experiment and the detection of particle
$1$ are assumed to be random and spacelike separated from those of
particle $2$.

As can be easily checked, in any EPR-type local realistic theory,
${\beta_{\rm EPR}} = 2^N$, while the value predicted by quantum
mechanics is $\beta_{\rm QM} = 4^N$, which violates the EPR bound by
an amount which grows as ${\beta_{\rm QM}}/{\beta_{\rm EPR}}=2^{N}$,
assuming perfect states and measurements. The remarkable point is
that this ``exponentially growing with size nonlocality'' can be
demonstrated by actual experiments if we use two-particle
hyperentangled states. In practice, we do not have perfect
correlations but
\begin{equation}
\langle X_1^{(1)} x_1^{(1)} \ldots
X_1^{(N)} x_1^{(N)} Y_2^{(1)} y_2^{(1)} \ldots Y_2^{(N)} y_2^{(N)}
\rangle = 1 - \epsilon,
\end{equation}
where $\epsilon \approx 0.15$ \cite{BDMVC06}.
In a worst-case scenario, each of the terms in $\hat{\beta}$
is affected by a similar error. Since the number of terms in
$\hat{\beta}$ is $4^N$, then we should take into account that our
value for $\beta_{\rm EPR}$ could be increased to
\begin{equation}
\beta_{\rm EPR}' \approx 2^N + 4^N 0.15.
\label{BetaEPRreal}
\end{equation}
Also, we must take into account the imperfection in
the preparation of the state which, in practice, is not
$|\Psi\rangle$, but $\rho = p |\Psi\rangle\langle \Psi| + (1-p)
\rho'$, with $p \approx 0.98$, and the specific form of the term
$\rho'$ depends on the physical procedure used to prepare and
distribute the state. Therefore, the expected experimental value of
$\hat{\beta}$ is
\begin{equation} \beta_{\rm QM}' \approx 0.98
\times 4^N + 0.02,
\label{BetaQMreal}
\end{equation}
The interesting point is that $\beta_{\rm QM}'$ still provides a
significant violation of the inequality $\hat{\beta} \le \beta_{\rm
EPR}'$, violation which exhibits a growing with size nonlocality. An
experiment for observing this effect for lower values of $N$ is
feasible using currently available capabilities \cite{BDMVC06}.




{\em Loophole-free Bell experiments.}---Experiments to test CHSH-BIs
have fallen within quantum mechanics and, under certain additional
assumptions, seem to exclude local realistic theories
\cite{Aspect99}. A particularly relevant loophole is the so-called
detection loophole \cite{Pearle70}. It arises from the fact that, in
most experiments, only a small subset of all the created pairs are
actually detected, so we need to assume that the detected pairs are
a fair sample of the created pairs. Otherwise, it is possible to
build a local model reproducing the experimental results. Closing
the detection loophole using the CHSH inequality requires $\eta \ge
0.83$ \cite{CH74,GM87}. The best of currently available
photodetection efficiency is $\eta = 0.33$ \cite{Kwiat05}.
There are several proposals for loophole-free experiments
\cite{KESC94}.
Garg and Mermin suggested that ``It is possible that $n \times n$
experiments with $n$ larger than $2$ can refute local realism with
lower detector efficiencies'' \cite{GM87}. Nevertheless, this
conjecture has not been proven to be true with known bipartite CHSH-BIs.
However, this effect can be observed by using EPR-BIs.

Consider the
previous EPR-BI for the state $|\Psi\rangle$. Let us calculate
the minimum detection efficiency required for a loophole-free
test.
If ${\cal N}({\cal AB}=1)$ is the number of pairs in which the
product of the results of measuring, for instance, ${\cal A} = X_1^{(1)}
x_1^{(1)} \ldots X_1^{(N)} x_1^{(N)}$ on particle $1$ and ${\cal B} = Y_2^{(1)}
y_2^{(1)} \ldots Y_2^{(N)} y_2^{(N)}$ on particle $2$ is $1$, and
${\cal N}$ is the total number of emitted pairs, then the
corresponding correlation is $\langle {\cal AB} \rangle =\left[{\cal
N}({\cal AB}=1)-{\cal N}({\cal AB}=-1)\right]/{\cal N}$. If $\eta$
is the detection efficiency of each and every one of the $4^N$ particle
detectors behind each analyzer, then the number of {\em detected}
pairs in which the product of the results of measuring ${\cal A}$ on particle $1$ and ${\cal B}$ on particle $2$ is $\pm 1$, is
related with the theoretical number by
\begin{equation}
{\cal N}_{\rm EXP} ({\cal AB}=\pm 1) = \eta^2 {\cal N}({\cal AB}= \pm 1).
\label{eqcero}
\end{equation}
On the other hand,
\begin{eqnarray}
{\cal N} & = & {\cal N}_{\rm EXP} ({\cal AB}=1) + {\cal N}_{\rm EXP}
({\cal AB}=-1) \nonumber
\\ & & + {\cal N}_{\rm EXP} ({\cal A}=\pm 1, {\cal B}=0)+ {\cal N}_{\rm EXP}
({\cal A}=0, {\cal B}=\pm 1) \nonumber
\\ & & + {\cal N}_{\rm EXP} ({\cal A}=0, {\cal B}=0),
\end{eqnarray}
where ${\cal N}_{\rm EXP} ({\cal A}=\pm 1, {\cal B}=0)$ is the number of
pairs in which when ${\cal A}$
is measured
on particle $1$ and ${\cal B}$
is measured
on particle $2$, and one detector corresponding to particle $1$ is activated,
but no detector corresponding to particle $2$ is activated.
We usually do not know ${\cal N}_{\rm EXP} ({\cal A}=0, {\cal B}=0)$,
because we cannot know ${\cal N}$; however, the relation between them is
\begin{equation}
{\cal N}_{\rm EXP} ({\cal A}=0, {\cal B}=0) = (1-\eta)^2 {\cal N}.
\label{eqtres}
\end{equation}
The probability that two or more detectors corresponding to the same
particle are activated simultaneously is assumed to be negligible.
What we obtain in an experiment is
\begin{eqnarray}
\langle {\cal
AB} \rangle_{\rm EXP} & = & \left[{\cal N}_{\rm EXP}({\cal
AB}=1)-{\cal N}_{\rm EXP}({\cal AB}=-1)\right] \nonumber \\ & &
\times \left[{\cal N}-{\cal N}_{\rm EXP} ({\cal A}=0, {\cal
B}=0)\right]^{-1}.
\label{eqdos}
\end{eqnarray}
Therefore, substituting (\ref{eqcero}) and (\ref{eqtres}) in
(\ref{eqdos}), we obtain
\begin{equation}
\langle {\cal AB} \rangle
= \frac{\eta^2}{1-(1-\eta)^2 }\langle {\cal AB} \rangle_{\rm EXP}.
\end{equation}
Therefore, taking into account the detection
efficiencies, the EPR-BI becomes
\begin{equation}
\frac{\eta^2}{1-(1-\eta)^2} \beta_{\rm QM} \le \beta_{\rm EPR},
\end{equation}
where $\beta_{\rm QM}$ and $\beta_{\rm EPR}$ must be replaced by
(\ref{BetaQMreal}) and (\ref{BetaEPRreal}), if we take the errors in
the state and the measurements into account. The remarkable point is
that the minimum $\eta$ required for a loophole-free test is a
function of the ratio $\beta_{\rm QM}/\beta_{\rm EPR}$. For
instance, assuming perfect states, for $N=1$ we recover the value
$\eta \approx 0.83$ \cite{CH74,GM87}. More interestingly, if we ask
which is the value of $N$ in order to get a loophole-free test
assuming the best currently available efficiency $\eta = 0.33$
\cite{Kwiat05}, the answer turns out to be $N \ge 6$ (i.e., 12
qubits per photon) taking into account the errors in the state and
the measurements. Alternatively, we can use higher dimensional
two-photon entangled states produced by fiber interferometers. This
has the advantage that each photon can be sent through a different
fiber and thus the local measurements can be spacelike separated.
The difficulty of having lower photodetection efficiencies at
telecom wavelengths could be compensated by the possibility of
producing entangled states in arbitrary high dimension
\cite{Gisin06}. Therefore, this approach could close the detection
loophole using currently available photodetection efficiencies.




The author thanks M. Barbieri, H.R. Brown, J.I. Cirac, F. De Martini, N. Gisin,
P.G. Kwiat, P. Mataloni, and M. \.{Z}ukowski for
useful discussions, and acknowledges support from Project
No.~FIS2005-07689.



\end{document}